\documentstyle[aps]{revtex}
\begin{document}
\title{Photonic production of $P$-wave states of $B_c$ mesons}
\author{A.V.~Berezhnoy, V.V.~Kiselev, A.K.~Likhoded\\
{\it Institute for High Energy Physics, Protvino 142284, Russia\\
E-mail: LIKHODED@MX.IHEP.SU}}
\maketitle
\begin{abstract}
Numerical calculations for the production of $P$-wave levels of $B_c$
quarkonium in $\gamma\gamma$ collisions are performed in the leading
$O(\alpha_s^2\alpha_{em}^2)$ order of perturbation theory. The total
cross-section of $P$-wave state production is about 10 \% of that for
the $S$-wave levels. The contribution of fragmentation
component ($6+6$ diagrams) is low, and the basic contribution is
determined by the recombination mechanism ($8$ Feynman diagrams).
The gauge invariant term of the $\bar b\to B_c$
fragmentation ($6$ diagrams) quite accurately reproduces the
result of the fragmentation model,
whereas there is a strong deviation of the $c\to B_c$ fragmentation term
from the predictions of the fragmentation model.
\end{abstract}
\section{Introduction}
At present, the production mechanism for the heavy quarkonium of mixed
flavour, $B_c$ meson [1], is quite reliably predicted theoretically. The
production of two pairs of heavy quarks  $b \bar b c \bar c$ is calculated in
the framework of perturbative QCD, whereas the hadronization of the
color-singlet $\bar b c$ pair into the meson is described in a
nonrelativistic potential model.

In $e^+e^-$ annihilation, the analysis of leading order approximation
in the perturbation theory of QCD at $M^2/s \ll 1$ allows one
to derive analytical expressions for differential cross-sections of the
$B_c$ meson production being treated as the process of the
$\bar b\rightarrow B_c$
fragmentation [2-5]. In the framework of the fragmentation mechanism,
the $S$ -wave level production dominates, and the $P$-wave state yield
is only about 10 \% with respect to the total number of produced mesons of
the $(\bar b c)$ family. However, as it was  found in photon-photon
and gluon-gluon collisions [6,7], there is  an additional contribution of
the recombination for the $S$-wave states. Such a term changes spectra as
well as the relative yields of pseudoscalar and vector states. The
recombination mechanism can result in an enhancement of the $P$-wave
level yield.

In this paper we consider the exact calculation of complete set of
the leading order diagrams in perturbation theory for the production of
the $(\bar b c)$ system $P$-wave states in $\gamma \gamma$ collisions and
make a comparison of the results with the
fragmentation model approximation.

\section{Calculation technique}

The $A^{SJj_z}$ amplitude of the $B_c$ meson production can be
expressed through
the amplitude of four free quarks production $T^{Ss_z}(p_i,k({\bf q}))$
and the orbital wave function of the $B_c$ meson, $\Psi^{Ll_z}({\bf q})$,
in the meson rest frame as
\begin{equation}
A^{SJj_z}=\int T^{Ss_z}(p_i,k({\bf q}))\cdot
\left (\Psi^{Ll_z}({\bf q}) \right )^* \cdot
C^{Jj_z}_{s_zl_z} \frac{d^3 {\bf q}}{{(2\pi)}^3},
\label{int}
\end{equation}
where $J$ and $j_z$ are the total spin of the meson and its projection on $z$
axis in the $B_c$ rest frame, correspondingly; $L$ and $l_z$ are the orbital
momentum and its projection; $S$ and $s_z$ are the sum of quark spins and its
projection; $C^{Jj_z}_{s_zl_z}$ are the Clebsh-Gordon coefficients;
$p_i$ are four-momenta of $B_c$, $b$ and $\bar c$,
${\bf q}$ is the three-momentum of $\bar b$ quark in the $B_c$ meson rest
frame; $k({\bf q})$ is the four-momentum, obtained from the
four-momentum $(0,{\bf q})$ by the Lorentz transformation from the
$B_c$ rest frame to the system, where the calculation of
$T^{Ss_z}(p_i,k({\bf q}))$ is performed. Then, the four-momenta
of $\bar b$ and $c$ quarks, composing the $B_c$ meson, will be determined
by the following formulae with the accuracy up to $|{\bf q}|^2$ terms
\begin{equation}
\begin{array}{c}
p_{\bar b}=\frac{m_b}{M}P_{B_c}+k({\bf q}), \\
p_{c}=\frac{m_c}{M}P_{B_c}-k({\bf q}),
\end{array}
\label{mom}
\end{equation}
where $m_b$ and $m_c$ are the quark masses, $M=m_b+m_c$,
and $P_{B_c}$ is the $B_c$ momentum. Let us  note that for the
$P$-wave states it is
enough to take into account only terms, linear over ${\bf q}$ in
eq.(\ref{int}).

The product of spinors $v_{\bar b} \bar u_c$, corresponding to the
$\bar b$ and $c$ quarks in the $T^{Ss_z}(p_i,k({\bf q}))$ amplitude of
eq.(\ref{int}), should be substituted by the projection operator
\begin{equation}
{\cal P} (\Gamma )=\sqrt{M} \left (\frac{\frac{m_b}{M} \hat P_{B_c}
+\hat k-m_b}{2m_b}
\right ) \Gamma \left (\frac{\frac{m_c}{M}\hat P_{B_c}-\hat k+m_c}{2m_c}
\right ),
\end{equation}
where $\Gamma=\gamma^5$ for $S=0$, or $\Gamma=\hat \varepsilon^*(P_{B_c},s_z)$
for $S=1$, where $\varepsilon(P_{B_c},s_z)$ is the polarization vector for the
spin-triplet state.

For the sake of convenience, one can express the ${\cal P}(\Gamma )$ operator
through the spinors of the following form
\begin{equation}
\begin{array}{c}
v_b'(p_b+k,\pm)=\left ( 1-\frac{\hat k}{2m_b} \right )v_b(p_b,\pm),\\
u_c'(p_c-k,\pm)=\left ( 1-\frac{\hat k}{2m_c} \right )u_c(p_c,\pm),
\end{array}
\label{spin}
\end{equation}
where $v_b(p_b,\pm)$ and $u_c(p_c,\pm)$ are the spinors with the given
projection of quark spin on $z$ axis in the $B_c$ meson rest frame.
Note, that the spinors in eq.(\ref{spin}) satisfy the Dirac equation
for the antiquark with the momentum $p_b+k$ and mass $m_b$ or for the
quark with the momentum $p_c-k$ and mass $m_c$ up to the linear order over
$k$ ( i.e. over ${\bf q}$, too), correspondingly.

One can easily show that the following equalities take place
\begin{equation}
\begin{array}{c}
\sqrt{\frac{2M}{2m_b2m_c}}\frac{1}{\sqrt{2}}\{v_b'(p_b+k,+)\bar u_c'(p_c-k,+)-
v_b'(p_b+k,-)\bar u_c'(p_c-k,-)\}=\\
={\cal P}(\gamma^5)+O(k^2),\\ \\

\sqrt{\frac{2M}{2m_b2m_c}}v_b'(p_b+k,+)\bar
u_c'(p_c-k,-)=\\
={\cal P}(\hat \varepsilon^*(P,-1))+O(k^2),\\ \\

\sqrt{\frac{2M}{2m_b2m_c}}\frac{1}{\sqrt{2}}\{v_b'(p_b+k,+)\bar u_c'(p_c-k,+)+
v_b'(p_b+k,-)\bar u_c'(p_c-k,-)\}=\\
={\cal P}(\hat \varepsilon^*(P,0))+O(k^2),\\ \\

\sqrt{\frac{2M}{2m_b2m_c}}v_b'(p_b+k,-)\bar
u_c'(p_c-k,+)=\\
={\cal P}(\hat \varepsilon^*(P,+1))+O(k^2).
\end{array}
\end{equation}
In the $B_c$ rest frame, the polarization vectors of the spin-triplet
state have the form
\begin{equation}
\begin{array}{l}
\varepsilon^{rf}(-1)=\frac{1}{\sqrt{2}}(0,1,-i,0), \\
\varepsilon^{rf}(0)=(0,0,0,1), \\
\varepsilon^{rf}(+1)=-\frac{1}{\sqrt{2}}(0,1,i,0).
\end{array}
\end{equation}
In calculations the Dirac representation of $\gamma$-matrices is used and
the following explicit form of the spinors is applied
\begin{equation}
\begin{array}{cc}
u(p,+)={ \frac{1}{\sqrt{E+m}}
\left (\begin{array}{c}
E+m\\
0\\
p_z\\
p_x+ip_y
\end{array}
\right ) },
&
u(p,-)={ \frac{1}{\sqrt{E+m}}
\left (\begin{array}{c}
0\\
E+m\\
p_x-ip_y\\
-p_z
\end{array}
\right ) }
\\
&\\
v(p,+)={ -\frac{1}{\sqrt{E+m}}
\left (\begin{array}{c}
p_z\\
p_x+ip_y\\
0\\
E+m
\end{array}
\right ) },
&
v(p,-)={ \frac{1}{\sqrt{E+m}}
\left (\begin{array}{c}
p_x-ip_y\\
p_z\\
0\\
E+m
\end{array}
\right ) }
\end{array}
\end{equation}
For the $P$-wave states in eq.(\ref{int}), the
$T^{Ss_z}\left (p_i,k({\bf q})\right )$ amplitude can be expanded
into the Taylor series up to the terms linear over ${\bf q}$.
Then one gets
\begin{equation}
A^{SJj_z}=iR_P'(0)\sqrt{\frac{2M}{2m_b2m_c}}
\sqrt{\frac{3}{4\pi}}C^{Jj_z}_{s_zl_z}
{\cal L}^{l_z}\left (T^{Ss_z}\left (p_i,k({\bf q})\right )\right ),
\label{main}
\end{equation}
where $R_P'(0)$ is the first derivative of the radial wave function at the
origin, and ${\cal L}^{l_z}$ has the following form
\begin{equation}
\begin{array}{l}
{\cal L}^{-1}=\frac{1}{\sqrt{2}}\left (\frac{\partial}{\partial q_x}
+i\frac{\partial}{\partial q_y} \right ), \\
{\cal L}^0=\frac{\partial}{\partial q_z}, \\
{\cal L}^{+1}=-\frac{1}{\sqrt{2}}\left (\frac{\partial}{\partial q_x}
-i\frac{\partial}{\partial q_y} \right ),
\end{array}
\label{dif}
\end{equation}
where $\frac{\partial}{\partial q_x}$,
$\frac{\partial}{\partial q_y}$, $\frac{\partial}{\partial q_z}$ are the
differential operators acting on $T^{Ss_z}\left (p_i,k({\bf q})\right )$ as
the function of ${\bf q}=(q_x,q_y,q_z)$ at ${\bf q}=0$.

As all considered matrix elements are calculated in the system
distinct from the $B_c$ rest frame, the four-momentum $k({\bf q})$
has been calculated by the following formulae
\begin{equation}
\begin{array}{l}
k^0=\frac{{\bf  v} \cdot {\bf q}}{\sqrt{1- {\bf v}^2}}, \\
{\bf k}={\bf q} +(\frac{1}{\sqrt{1- {\bf v}^2}}-1)\frac{{\bf  v}
\cdot {\bf q}}
{{\bf v}^2}{\bf  v},
\end{array}
\label{Lor}
\end{equation}
where ${\bf  v}$ is the $B_c$ velocity in the system, where the calculations
are performed. The matrix element
$T^{Ss_z}\left (p_i,k({\bf q})\right )$ is computed, so that the
four-momenta of $\bar b$ and $c$ quarks are determined by eq.(\ref{mom}),
taking into account eq.(\ref{Lor}).

The first derivatives in eq.(\ref{dif}) are substituted by the
following approximations
\begin{equation}
\frac{\partial T^{Ss_z}\left (p_i,k({\bf q})\right )}{\partial q_j}
\vert_{{\bf q=0}}\approx \frac{T^{Ss_z}\left (p_i,k({\bf q}^j)\right )
-T^{Ss_z}\left (p_i,0\right )}{\triangle},
\end{equation}
where $\triangle$ is  some small value, and ${\bf q}^j$ have the following form
\begin{equation}
\begin{array}{c}
{\bf q}^x=(\triangle,0,0),\\
{\bf q}^y=(0,\triangle,0),\\
{\bf q}^z=(0,0,\triangle).
\end{array}
\end{equation}
With the chosen values of quark masses and interaction energies, the
increment value $\triangle=10^{-5}$ GeV has provided the stability of
4-5 meaning digits in the  squared matrix elements summed over $j_z$
for all $P$-wave states with the given value of $J$ and $S$,
when one has performed the Lorentz transformations along the beam axis or
the rotation around the same axis.

One has to note that because of such transformations,
the new vectors $k({\bf q}^j)$
do not correspond to the transformed old vectors. Therefore, the applied test
is not only a check of the correct typing of the
$T^{Ss_z}\left (p_i,k({\bf q})\right )$ amplitude, but it is also the check
of correct choice of the phases in eq.(\ref{main}).

The  matrix element $A^{SJj_z}$ squared,  which is calculated by the method
described above, must be summed over $j_z$ as well as the spin states of free
$b$ and $\bar c$ quarks. It also must be  averaged over spin projections
of initial particles.

The phase space integration has been made by the Monte Carlo method of RAMBO
program [8].

\section{Discussion of results}

To check the numerical way of the amplitude calculation for the $P$-wave
level production of $B_c$, we have considered the production of
these states in $e^+e^-$ annihilation, where the analytical expressions
for the differential cross-sections were derived in the $M^2/s \ll 1$ limit
[5]. Those expressions define the functions of the
$\bar b \rightarrow B_c (L=1)$
fragmentation. It should be noted that in the accurate consideration
of the fragmentation mechanism for the $B_c (L=1)$ production in
$e^+e^- \rightarrow \gamma^* \rightarrow  B_c+X$, one can see that in
addition to the $\bar b$ fragmentation one has to account for the
$c$ quark fragmentation into $B_c$. Moreover, the most significant role
is played by the $c$ fragmentation into $^3 P_0$ state. Therefore, to
compare with the numerical results we use the
analytical expressions  accounting for both $\bar b$ and $c$
fragmentation. As one can see in Fig.1, the distributions of
$^1P_1$, $^3P_0$, $^3P_1$, $^3P_2$ level
\footnote{In the $\bar b c$ system the quark spin-dependent corrections to
the  heavy quarkonium potential lead to the mixing of $^3P_1$ and $^1P_1$
levels [9]. In the fragmentation model this  mixing results in
redefinition of corresponding fragmentation functions,
i.e. in the introduction
of additional functions for the $1^+$ and ${1^+}'$ states [5]. However,
first, the mixing effect does not influence  the consideration of the
physical mechanism of photonic
production of the $P$-wave states, and second, in the studied approach
the $^SP_J$-state masses are degenerated over $J$, so that the isolation
of the $S=0$ and $S=1$ components in the $J=1$ state is rather conventional.
That is why we restrict ourselves by the consideration of distributions
for the $^SP_J$ states.}
production,
$d\sigma /dz$ ($z=2|{\bf P}_{B_c}|/\sqrt{s}$ with
${\bf P}_{B_c}$ being the three-momentum of $B_c$ meson in the c.m.s.),
calculated numerically and
given analytically, coincide with each other at the same set of
parameters, which are give below
\begin{equation}
\begin{array}{l}
\alpha_{em}=1/128,\\
\alpha_s=0.2,\\
m_b=5.0\  {\rm GeV},\\
m_c=1.7\  {\rm GeV},\\
|R_P'(0)|^2=0.201\  {\rm GeV}^5.
\end{array}
\end{equation}
Thus, the performed verification convinces us that the calculation method
used is quite accurate.

Let us  consider the $P$-wave level production of $B_c$ in
$\gamma \gamma$ collisions.
 The cross-section of the $P$-wave level production at various energies of
interacting photons are presented in Tab.1 and Fig.2. One can see in Fig.2
that in the region of interest the energy dependence of the summed
cross-section
at the chosen  $m_b$ and $m_c$ values is quite accurately described by the
following approximation, shown as solid line in Fig.2,
\begin{equation}
\sigma_{B_c(L=1)}=130\cdot \left ( 1- \left ( \frac{2(m_b+m_c)}{\sqrt{s}}
\right )^2 \right )^{2.7}
\cdot \left ( \frac{2(m_b+m_c)}{\sqrt{s}} \right )^{1.32} {\rm fb}.
\label{emp}
\end{equation}
The summed fragmentation contribution, obtained as the product of the
$\gamma \gamma \rightarrow b \bar b$ and $\gamma \gamma \rightarrow c \bar c$
cross-sections and the corresponding probabilities of the fragmentation
($5.34 \cdot 10^{-5}$ for the $\bar b$ quark fragmentation and
$1.58 \cdot 10^{-6}$ for the $c$ quark one,
respectively), is shown in the same figure.
One can see in Fig.2 that the fragmentation contribution, evaluated in the
model, is small at high energies, where the application of this model could be
sound for the $\bar b$ fragmentation, at least. The fragmentation model
overestimates the exact result obtained over the complete set of leading
order diagrams at low energies close to $\sqrt{s} \le 30$ GeV. This
overestimation has a simple explanation and it is related with incorrect
evaluation of phase space, since in the fragmentation model, one uses
the two-particle phase space instead of the three-particle one in the
exact calculations. For the correct study of the fragmentation mechanism
the  $M^2/s\ll 1$ condition is necessary to be satisfied. Therefore,
in what follows, we will restrict ourselves by the consideration of
differential distributions at $\sqrt{s}=100$ GeV, where the mentioned
condition is certainly valid.

The $d\sigma /dz$ distribution for $^1 P_1$ state
production cross-section in $\gamma \gamma$ collisions
is shown as the solid line histogram in
Fig.3a. The dashed line histogram in the same figure denotes the gauge
invariant contribution of six diagrams, where the $c\bar c$ pair is
emitted from the $b$ or $\bar b$ quark line, i.e. the $\bar b$ quark
fragmentation into $B_c$ meson takes place. This exactly calculated term is
compared with the prediction of fragmentation model, considered in
[5] and presented as the smooth dashed curve. One can see that the exact
result for $\bar b \to B_c$ is quite
accurately described by analytical expression of
fragmentation model, whereas the $c$ fragmentation diagrams contribution,
shown as the dotted line histogram, is larger than the predictions of
fragmentation model, so that in absolute values, it is larger than the
$\bar b$ quark fragmentation term. Remember, that the analogous picture
takes place
also for the photonic production of $S$-wave states of $B_c$ meson [6].
One can see in Fig.3a, that as well as in the production of $S$-wave levels,
the fragmentation contribution does not dominate, and the main
contribution is determined by the recombination diagrams. However, for the
correct study of the fragmentation mechanism in photon-photon
collisions, one must consider the
spectra not over the total energy of the meson, but over its transverse
momentum, giving an additional scale of energy, so that one can expect
the factorization of the heavy quark hard production at large
$P_T \gg M_{B_c}$  and the forthcoming fragmentation, where the particle
virtualities are of the order of the quark masses.

The cross-section distributions over the transverse momentum of
$^1P_1$ level of the $B_c$ meson are shown in Fig.3b. The solid line
histogram denotes the result of calculations over the complete set of diagrams,
the dashed histogram and curve show the contribution by the $\bar b$
fragmentation
diagrams and the prediction of  fragmentation model, correspondingly,
and the
$c$ fragmentation diagram contribution (the dotted histogram) is
compared with the corresponding prediction of fragmentation model (the dotted
curve) at $\sqrt{s}=100$ GeV.

As well as in the $S$-wave state production, the fragmentation mechanism
does not dominate even at large $P_T$.

The analogous distributions for the $^3 P_0$, $^3 P_1$, $^3 P_2$ states are
presented in Figs.4a-6b. In Figs.5-6, where the $^3 P_1$, $^3 P_2$
spectra are shown, one can draw the conclusions, which repeat the statements
concerning the $^1 P_1$ production. The picture of the $^3 P_0$ meson
production slightly differs from the general case. As one can see in Fig.4b,
the $\bar b$ quark fragmentation at large transverse momenta plays a much
greater role than that in the production of other $P$-wave levels, while,
contrary, the diagrams corresponding to the $c$ fragmentation, are less
essential than in other cases.

\section{Conclusion}

In this paper we have performed the numerical calculation for the $P$-wave
level production of $B_c$ mesons in $\gamma \gamma$ collisions in the
leading $O(\alpha^2_s \alpha^2_{em})$ order of the perturbation theory. From
the theoretical point of view, the consideration of
$\gamma \gamma$  collisions in this respect is of special interest, since,
on the one hand, in the photonic production of $B_c$ as well as in
$e^+e^-$ annihilation, one can isolate the gauge invariant set of
fragmentational diagrams, and, on the other hand, there is also
the contribution of recombination type diagrams,
which are essential in the consideration of a
more complicated case of the hadronic $B_c$ production [7].

The performed calculations show  that:
\begin{enumerate}
\item
The total cross-section of $P$-wave state production of $B_c$ is about 10 \%
 in respect to the production of $S$-wave levels, as it takes place in
$e^+e^-$ annihilation.
\item
As well as in the $S$-wave level production, the recombination mechanism
dominates, and the fragmentation one is small.
\item
The $\bar b$ fragmentation diagram
contribution is quite accurately described by
the fragmentation function at large $P_T$ as well as low one, whereas
for the $c$ fragmentation diagrams, the fragmentational picture is broken
completely.
\end{enumerate}

This work is  supported, in part,  by the International Science Foundation
grants NJQ000 and NJQ300. The work of A.V.~Berezhnoy
has been made possible by a fellowship of INTAS Grant 93-2492 and  one
of International Soros Science Education Program Grant A1377 and
is carried out within the research program of International Center for
Fundamental Physics in Moscow.

\section*{References}
\begin{enumerate}
\item
 {\it Gershtein~S.S., Kiselev~V.V., Likhoded~A.K., Tkabladze~A.V.}//
Uspekhi Fiz. Nauk. 1995. V. 165. P. 3.
\item
 {\it Clavelli~L.}// Phys. Rev. 1982. V. D26. P. 1610;
 {\it Ji~C.-R. and Amiri~R.}// Phys. Rev. 1987. V. D35. P. 3318;
 {\it Chang~C.-H. and Chen~Y.-Q.}//Phys. Lett. 1992. V. B284 P. 127.
\item
{\it Braaten~E., Cheung~K., Yuan~T.C.}// Phys.Rev. 1993. V. D48. P. 4230.
\item
{\it Kiselev~V.V., Likhoded~A.K., Shevlyagin~M.V.}// Z.Phys. 1994.
V. C63. P. 77.
\item
{\it Yuan~T.C.}// Phys.Rev. 1994. V. D50. P. 5664.
\item
{\it Berezhnoy~A.V., Likhoded~A.K., Shevlyagin~M.V.}//
 Phys.Lett. 1995. V. B342. P. 351;
{\it Ko\l odzej~K., Leike~A., R\"uckl~R.}// Preprint MPI-PhT/94-84,
LMU-23-94, 1994.
\item
{\it Berezhnoy~A.V., Likhoded~A.K., Shevlyagin~M.V.}// Yad.Fiz. 1995. V. 58. P.
 730;
{\it Berezhnoy~A.V., Likhoded~A.K., Yushchenko~O.P.}//
 Preprint IHEP 95-59. Protvino, 1995; hep-ph/ 9504302;
{\it Ko\l odzej~K., Leike~A., R\"uckl~R.}// Preprint MPI-PhT/95-36, 1995;
hep-ph/9505298.
\item
{\it Kleiss~R., Stirling~W.J., Ellis~S.D.}// Comp. Phys. Commun. 1986.
V. 40. P. 356.
\item
{\it Gershtein~S.S. at al.}// Phys. Rev. 1995. V. D51. P. 3613.
\end{enumerate}

\begin{table}[p]
\begin{center}
\caption{The dependence of production cross-sections for different $P$-wave
states of $B_c$ on the total energy of colliding photons.
(The calculation errors are shown in parenthesis).}
\begin{tabular}{||c|c|c|c|c||}
\hline
     & & & &       \\
$\sqrt{s}$, GeV  & $\sigma_{^1P_1}$,  fb
 &  $\sigma_{^3P_0}$, fb
 &  $\sigma_{^3P_1}$, fb
 &  $\sigma_{^3P_2}$, fb  \\
     & & & &       \\
\hline
15.  & 1.064(2)  & 0.1329(3) & 0.08707(18) & 0.894(2)  \\
20.  & 5.703(11) & 0.968(2)  & 1.147(2)    & 8.098(15) \\
30.  & 7.56(3)   & 1.409(5)  & 2.84(1)     & 12.11(5)  \\
40.  & 6.72(4)   & 1.269(9)  & 3.063(18)   & 11.01(7)  \\
60.  & 4.68(5)   & 0.882(12) & 2.42(3)     & 7.7(1)    \\
80.  & 3.32(5)   & 0.624(15) & 1.80(3)     & 5.48(12)  \\
100. & 2.58(3)   & 0.493(9)  & 1.42(2)     & 4.32(7)   \\
\hline
\end{tabular}
\end{center}
\end{table}

\newpage
\section*{Figure captions}
\begin{itemize}
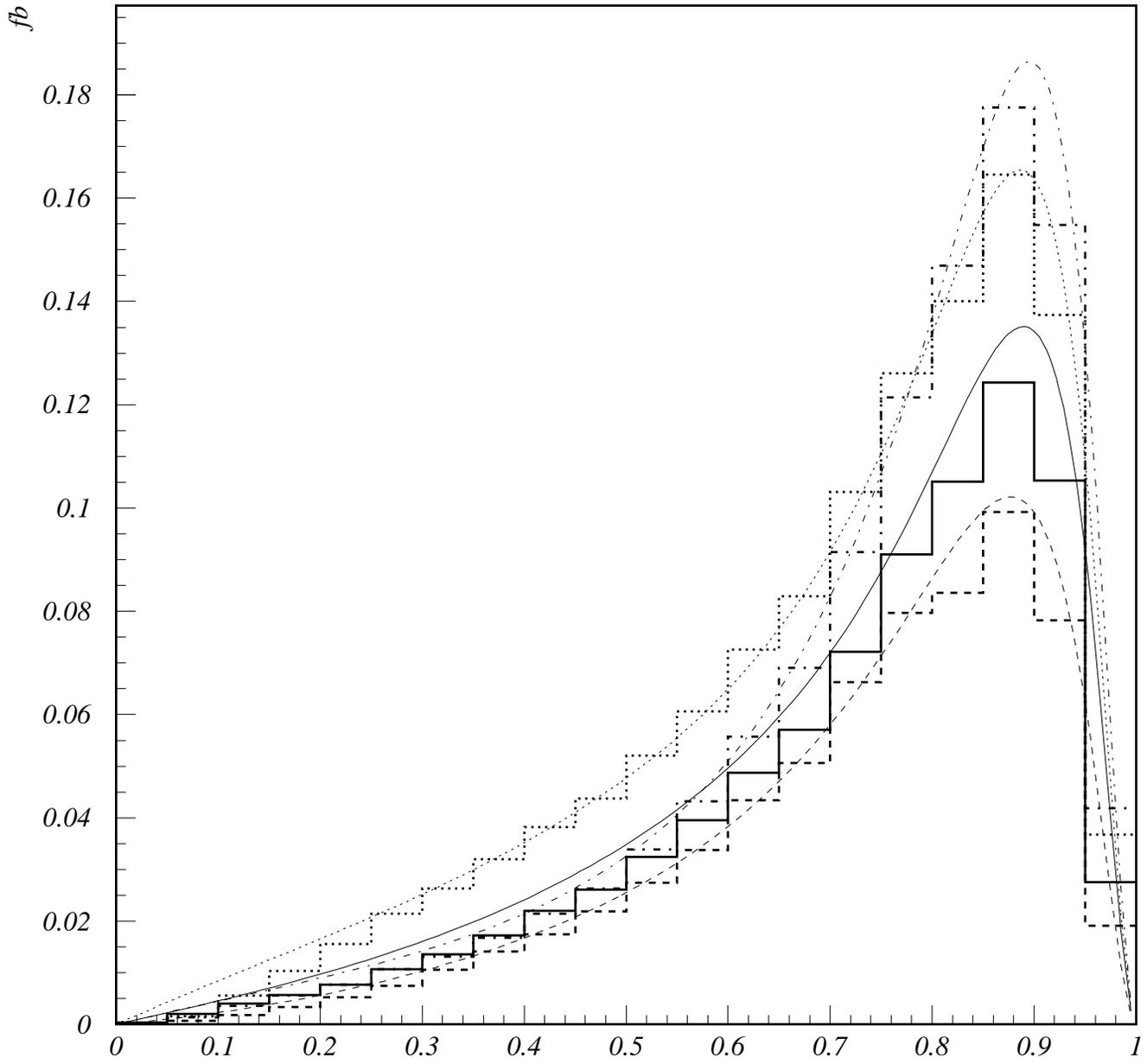
\item[Fig. 1.] The $B_c$ production cross-section
distributions over $z$ in the $e^+e^- \rightarrow \gamma^* \rightarrow  B_c+X$
process are presented as the histograms in comparison with the
fragmentation model predictions shown as the smooth curves, at $\sqrt{s}=100$
GeV for the following
$B_c$ states: $^1P_1$ (solid line),  $^3P_0$ (dashed line), $^3P_1$
(dotted line), $^3P_2$ (dotted-dashed line).
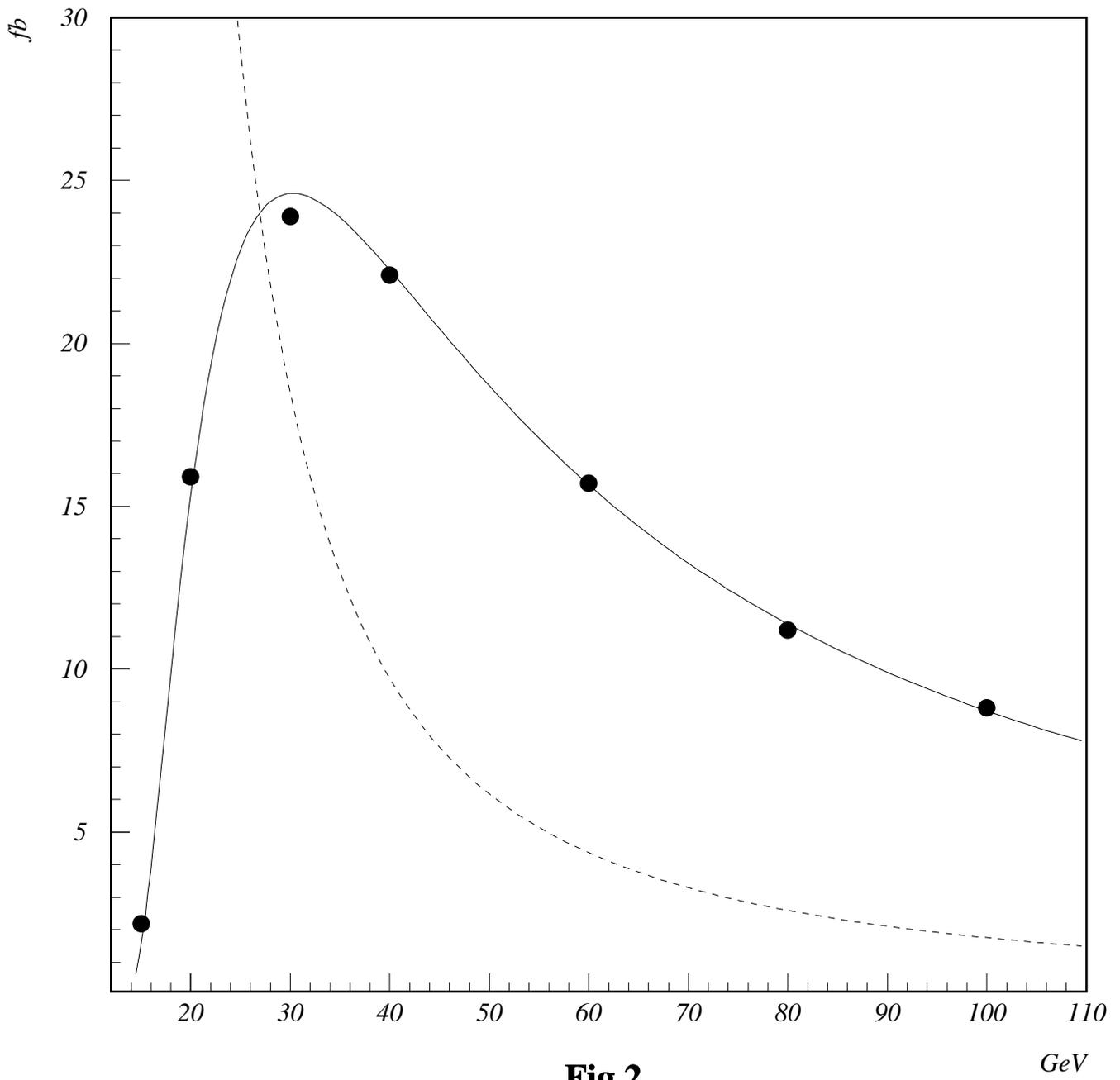
\item[Fig. 2.] The summed cross-section dependence of $P$-wave states
on the energy of interacting photons is marked by $(\bullet)$ in comparison
with the prediction of $\bar b$ and $c$ quark fragmentation model
 (dashed curve).
The solid line corresponds to the approximation of (\ref{emp}).
\item[Fig. 3.]
\begin{itemize}
\item[a.] The $z$ distributions, corresponding to the photonic production of
$^1P_1$ state at the interaction energy 100 GeV. The total
result is presented by the solid line histogram, the $b$ fragmentation diagrams
contribution (dashed histogram) is compared with the prediction of
fragmentation model (dashed curve), the $c$ fragmentation diagrams term
is denoted as the dotted histogram in comparison with the fragmentation
model result (dotted curve).
\item[b.] The transverse momentum distributions, corresponding to the
photonic production of $^1P_1$ state at the interaction energy 100 GeV.
The notations of different contributions are the same as in Fig. 3a.
\end{itemize}
\item[Fig. 4a,b.] The $^3P_0$ spectra, denoted as in Fig.3a,b.
\item[Fig. 5a,b.] The $^3P_1$ spectra, denoted as in Fig.3a,b.
\item[Fig. 6a,b.] The $^3P_2$ spectra, denoted as in Fig.3a,b.
\end{itemize}
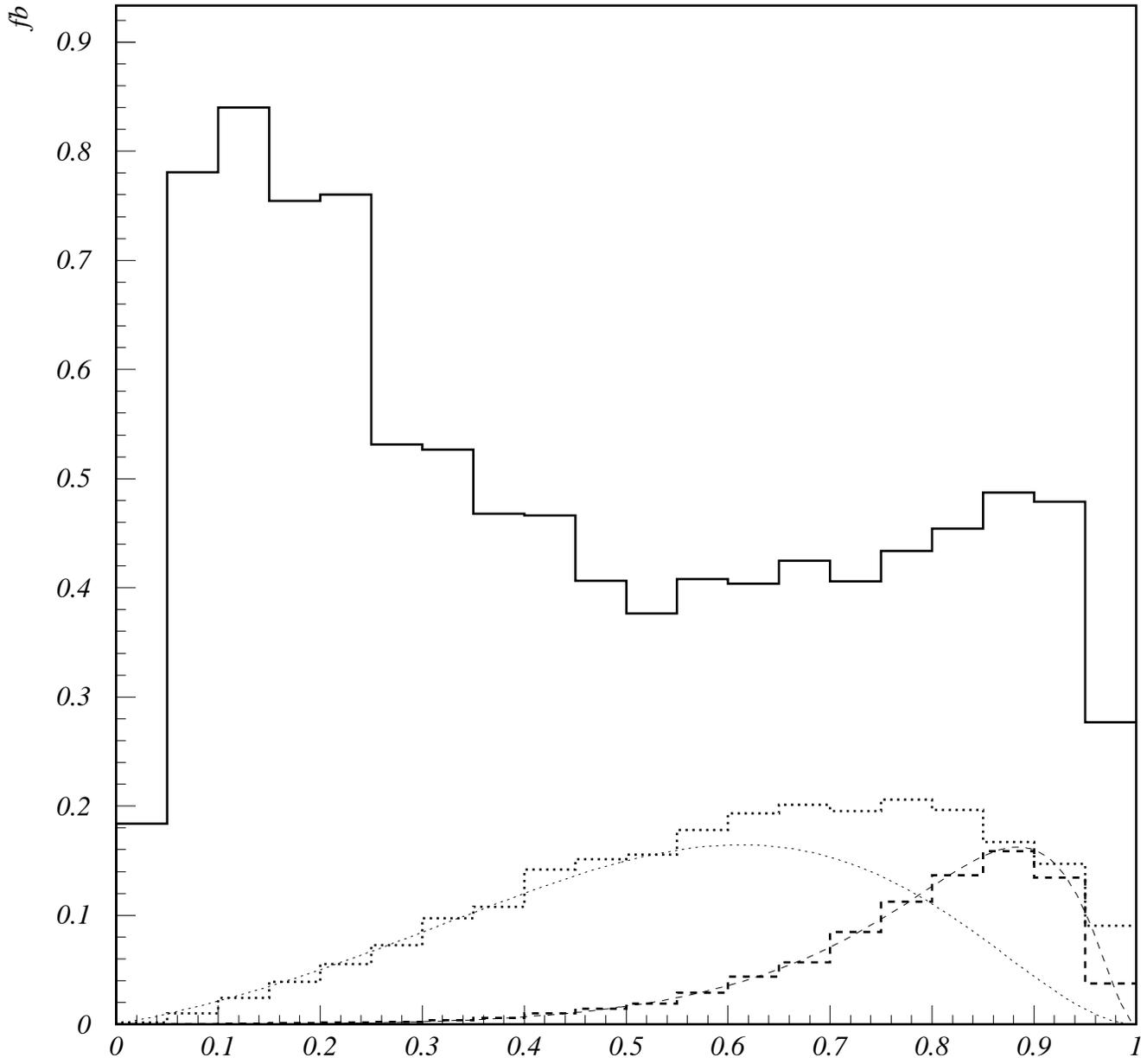
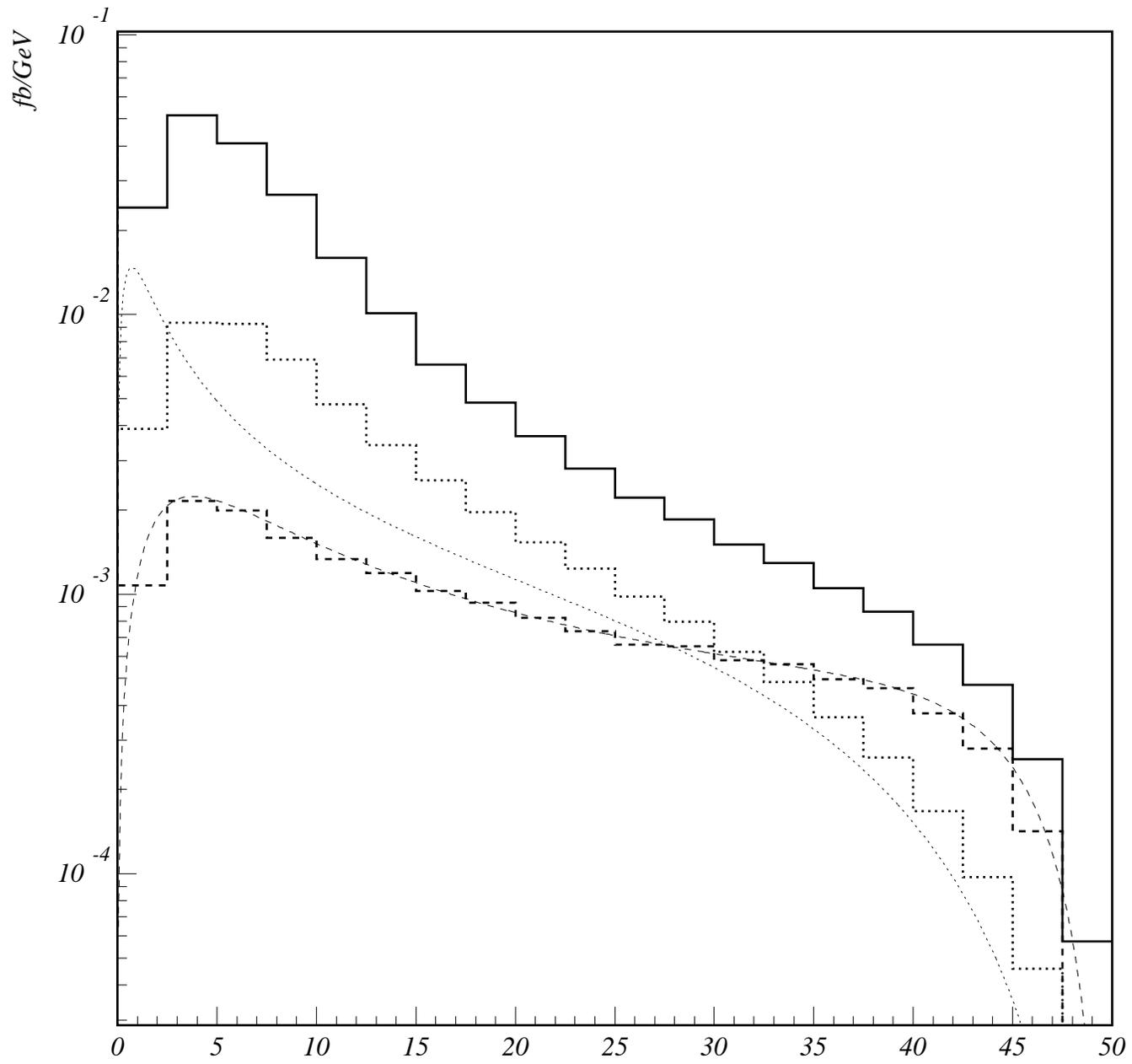
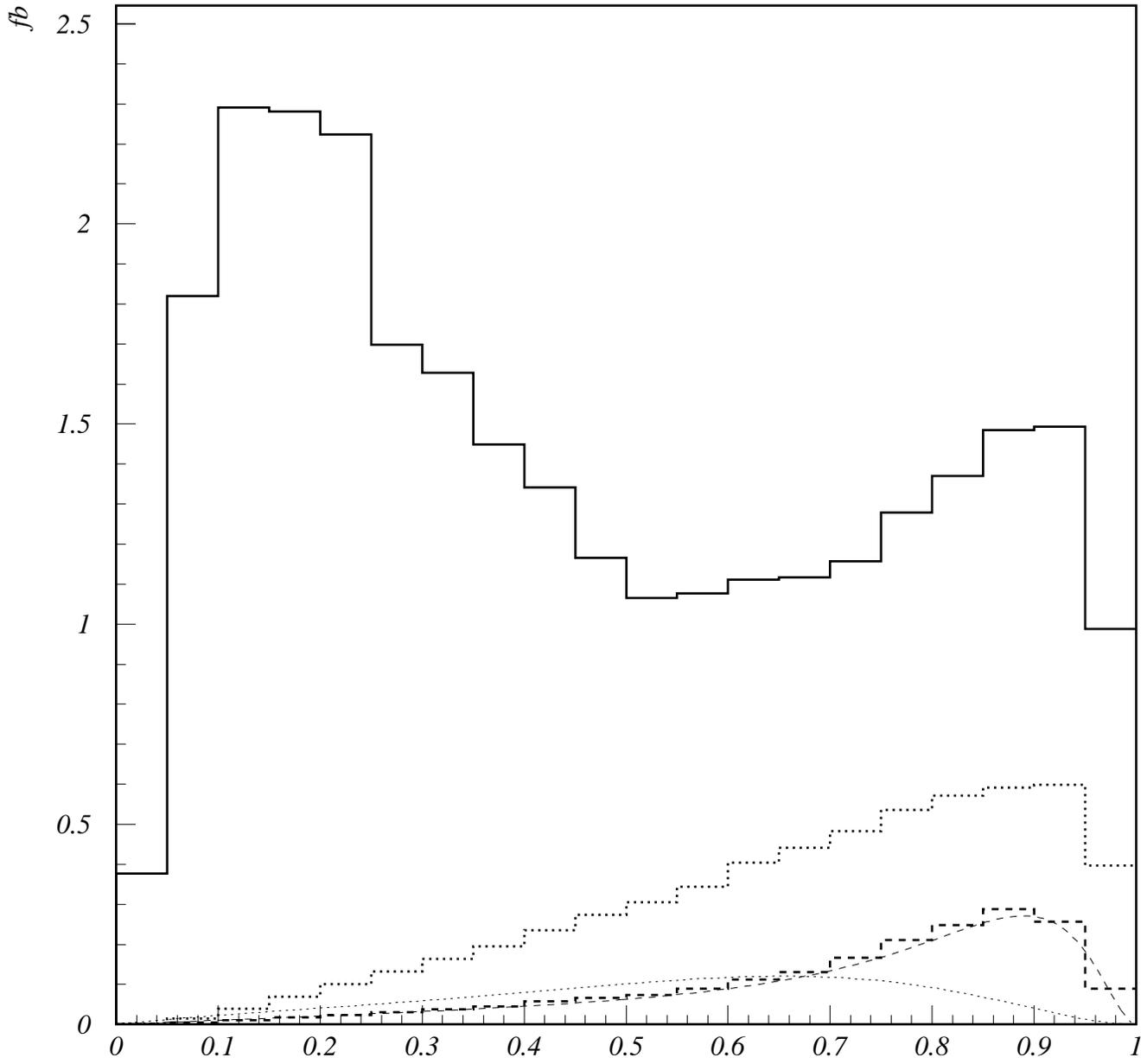
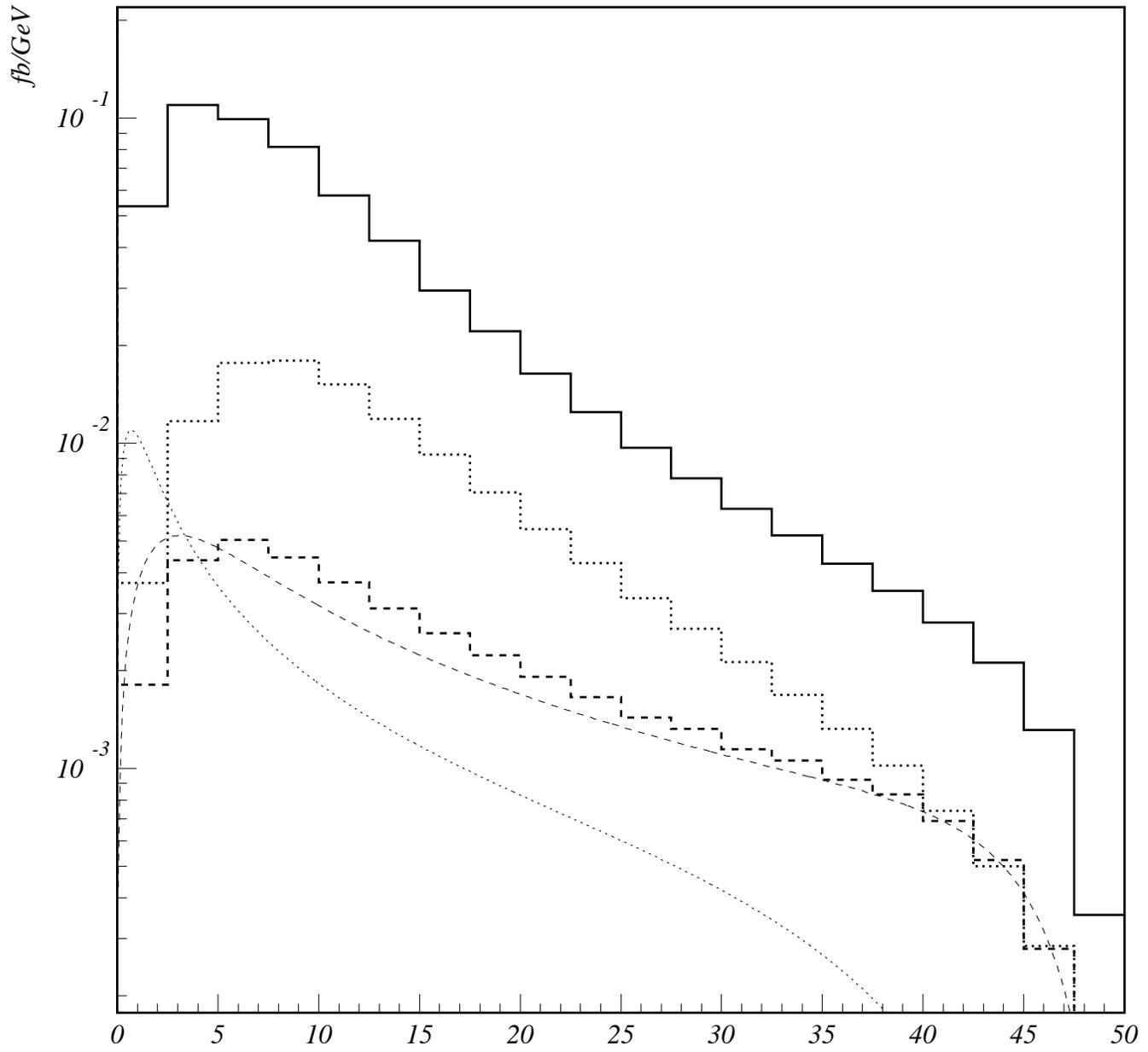
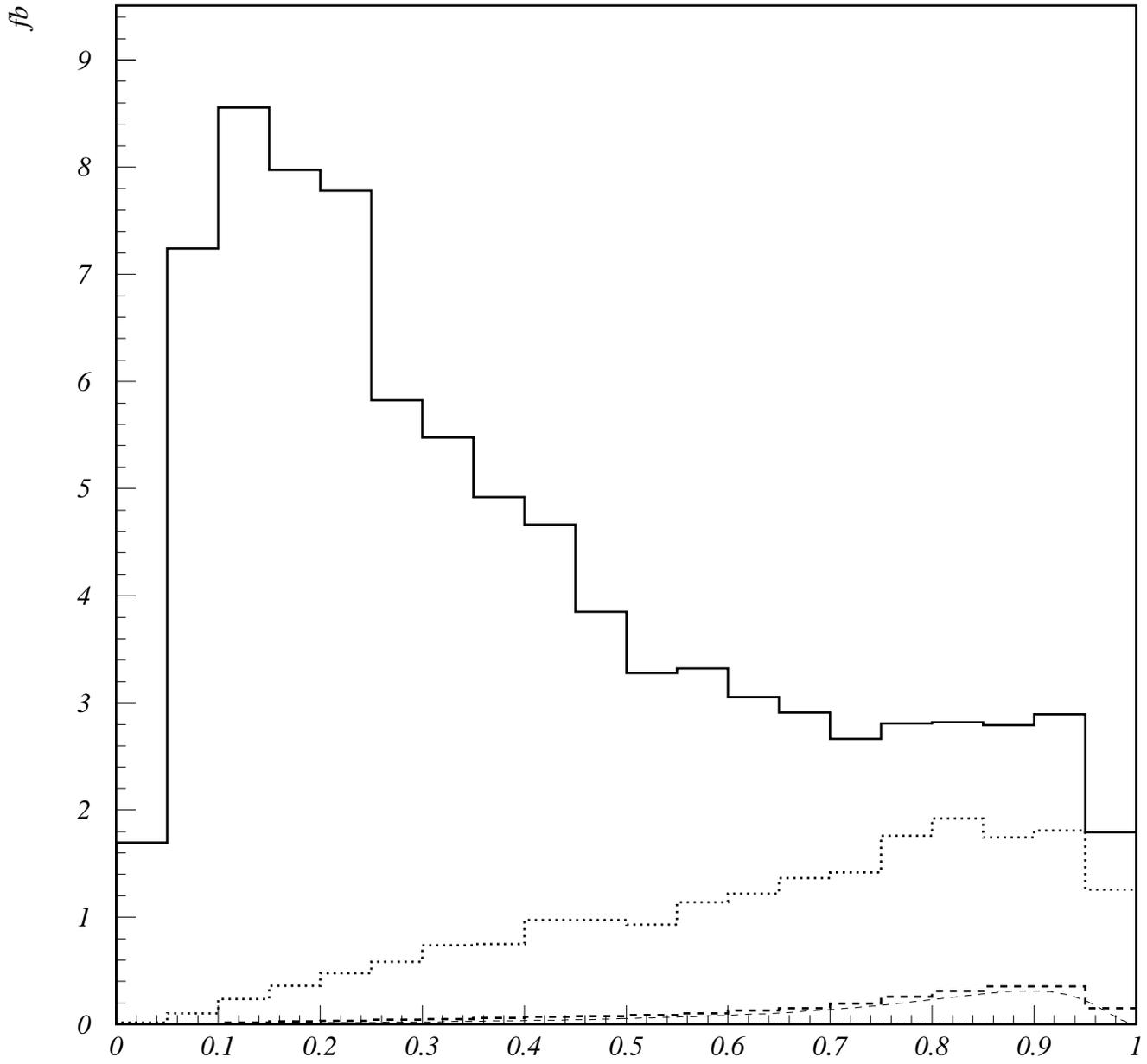
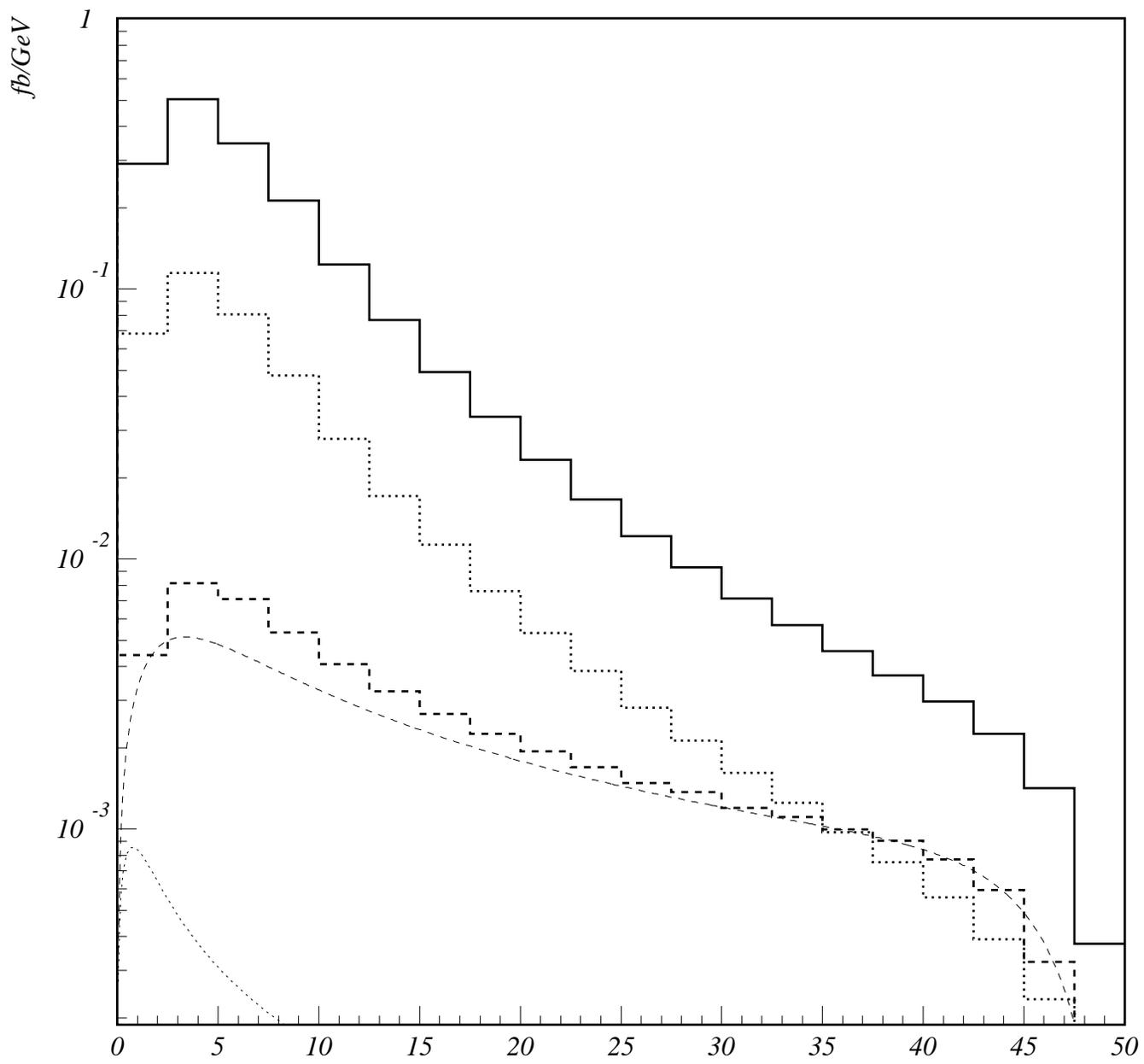
\end{document}